\documentclass[structabstract]{aa}  
\usepackage{graphicx}
\usepackage{txfonts}
\def\XMM{XMM-{\sl Newton}}
\def\suz{{\sl Suzaku}}
\def\ASCA{{\sl ASCA}}
\def\chan{{\sl Chandra}}
\def\FeKa{Fe K$\alpha$}
\def\FeKb{Fe K$\beta$}
\def\FeXXV{Fe\,{\sc xxv}}
\def\FeXXVI{Fe\,{\sc xxvi}}
\begin{document}
   \title{Probing the unified model in NGC 7314}

%   \subtitle{}

   \author{J. Ebrero\inst{1}
          \and
          E. Costantini\inst{1}
          \and
          J. S. Kaastra\inst{1,2}
          \and
          B. De Marco\inst{3,4}
          \and
          M. Dadina\inst{4}
          }

   \institute{SRON Netherlands Institute for Space Research,
              Sorbonnelaan 2, 3584 CA, Utrecht, The Netherlands\\
              \email{J.Ebrero.Carrero@sron.nl}
              \and
              Astronomical Institute, University of Utrecht,
              Postbus 80000, 3508 TA, Utrecht, The Netherlands
              \and
              Dipartimento di Astronomia, Universit\'a di Bologna, Via Ranzani 1, 40127, Bologna, Italy
              \and
              INAF - IASF Bologna, via Gobetti 101, 40129 Bologna, Italy
             }

   \date{Received <date>; accepted <date>}

% \abstract{}{}{}{}{} 
% 5 {} token are mandatory
 
  \abstract
   {}
   {We aim at studying the complex absorbed X-ray spectrum of the Narrow Line Seyfert 1 galaxy NGC 7314.}
   {We collected available public X-ray data from the archives of \XMM{}, \suz{}, and \ASCA{}. The spectra were analyzed using the fitting package {\sl SPEX}.}
   {We find evidence of intrinsic neutral and ionized absorption in the \XMM{} EPIC-pn spectrum. The ionized gas presents three significantly distinct ionization phases, although its kinematic properties could not be disentangled. At least two of these phases are also detected in the RGS spectrum, although with less significance due to the low statistics. The \ASCA{} and \suz{} spectra show larger neutral absorption but no ionized gas signatures. The \FeKa{} emission line is detected in all the observations and, additionally, \FeXXVI{} in the EPIC-pn spectrum, and \FeKb{} in the \suz{} XIS spectrum.}
   {Using this observational evidence we construct a consistent picture of the geometry of the system in the context of the unified model of active galactic nuclei. The different observational properties are thus interpreted as clouds of neutral gas moving across our line of sight, which would be grazing a clumpy dusty torus.}


   \authorrunning{J. Ebrero et al.}
   \titlerunning{Probing the unified model in NGC 7314.}

   \maketitle
%
%________________________________________________________________

\section{Introduction}
\label{intro}

X-rays play a key role in the study of active galactic nuclei (AGN, hereafter) since they originate in their innermost regions, thus allowing to investigate the environment and physical properties of the circumnuclear gas.

The differences between type-1 and type-2 AGN are ascribed to absorbing material along our line of sight. From an observational point of view this gas has the effect of obscuring the AGN emission from the inner regions (e.g. optical/UV spectral lines from the broad line region, BLR) up to X-ray energies. A Compton-thick molecular torus located at pc-scale distances from the nucleus is claimed to be responsible for this dichotomy (Antonucci \cite{Ant93}).

However, there is increasing evidence that such a simple scenario may not hold for all objects. In some sources weak or totally absent X-ray absorption along with the lack of optical broad lines have been reported (Panessa \& Bassani \cite{PB02}, Wolter et al. \cite{Wol05}, Bianchi et al. \cite{Bia07}). On the other hand, a number of Seyfert 2 galaxies showed significant variability in their absorbing column density $N_{\rm H}$ (Risaliti et al. \cite{Ris02}) on short time scales (from months up to few hours) thus indicating that the absorbing material must reside much closer to the nucleus than the putative torus, e.g. at the BLR length scale (NGC 4388, Elvis et al. \cite{Elvis04}; NGC 1365, Risaliti et al. \cite{Ris07}; NGC 4151, Puccetti et al. \cite{Puc07}).

Moreover, there is evidence of ionized gas in the soft X-ray spectrum of many type-1 AGN (e.g. Reynolds \cite{Rey97}; Piconcelli et al. \cite{Pic05}), the so-called warm absorber (WA). The possible origin and location of the WA in the context of the unification picture of AGN is a hot topic in modern extragalactic astronomy, as several possible hypothesis may explain the observed features, such as evaporation from the torus (Krolik \& Kriss \cite{KK01}), outflowing winds from the accretion disk (e.g. Elvis \cite{Elvis00}), or production in an ionization cone (see e.g. Bianchi et al. \cite{Bia06}). The study of Blustin et al. (\cite{Blu05}) on 23 AGN suggests that the WA in most of nearby Seyfert galaxies likely originates in outflows from the molecular torus.

Within this scenario it does result extremely difficult to detect warm and neutral absorbers simultaneously in a single source because the torus would therefore obscure the inner regions of the AGN, where the WA is likely located. Only a few detections have been reported so far, e.g. H0557-385 (Ashton et al. \cite{Ash06}; Longinotti et al. \cite{Lon09}), I Zw 1 (Costantini et al. \cite{Cos07}).

NGC 7314 is a nearby ($z=0.004763$, Koribalski et al. \cite{Kor04}) and bright ($F(2-10~{\rm keV})=3.7 \times 10^{-11}$~erg cm$^{-2}$~s$^{-1}$, this study) Seyfert galaxy cataloged in the Piccinotti sample (Piccinotti et al. \cite{Pic82}). This source has been optically classified as a Seyfert 2 (e.g. Morris \& Ward \cite{MW88}). A broad component of H$\alpha$ measured in a {\it Hubble} Space Telescope spectrum is consistent with a Seyfert 1.9 classification (Hughes et al. \cite{Hug03}). However, the source displays rather peculiar X-ray properties. Indeed its X-ray flux does experience extreme variability events ($\sim$50\% flux variations on time scales of a few thousands of seconds), while the measured X-ray spectral index appears unusually steep ($\Gamma \sim 2$). These properties, together with the relative narrowness of its H$\beta$ emission lines, are typical of narrow line Seyfert 1 galaxies (NLS1, e.g. Pounds et al. \cite{Pou95}; Boller et al. \cite{Bol96}). Thus NGC 7314 has been proposed to be a good type 2 counterpart of a NLS1 (Dewangan \& Griffiths \cite{DG05}). We note, however, that the sole width of the H$\beta$ line might not be enough for a source to be classified as a NLS1 (see e.g. Sulentic et al. \cite{Sul00}; V\'eron-Cetty \& V\'eron \cite{VCV06}).

The soft X-ray spectrum of NGC 7314 shows great complexity; the \XMM{} spectrum of the source in both the RGS and EPIC-pn reveals neutral absorption at the redshift of the source ($N_{\rm H}\sim 5 \times 10^{21}$~cm$^{-2}$) and hints of an intrinsic ionized absorber (Branduardi-Raymont et al. \cite{BR02}).

The Fe K region of the source was observed with \ASCA{} and \chan{} showing that the Fe K$\alpha$ line at 6.4~keV may present a broad component plus a narrow core. The study of the variability of the Fe K$\alpha$ line profile showed that the core of the line does not respond to the rapid variations in the continuum as fast as the wings. This suggests different origins for both components, the latter originating in distant matter rather than in the accretion disk (Yaqoob et al. \cite{Yaq96}, \cite{Yaq03}). NGC 7314 is therefore a complex source that deserves further study. The absorption properties in the soft X-ray spectrum throughout the different archival observations of this source are discussed in this work in the context of the unification model of AGN. The timing properties and the broad band spectrum of NGC 7314 will be discussed in a forthcoming paper (De Marco et al., in preparation).

This paper is organized as follows: in Sect.~\ref{Xdata} we describe the X-ray data used and how they were reduced, in Sect.~\ref{analysis} we present the results of the analysis of the various X-ray spectra, while in Sect.~\ref{discussion} we discuss the results. Finally, our conclusions are summarized in Sect.~\ref{conclusions}. In what follows we assume a cosmological framework with $H_0=70$~km s$^{-1}$ Mpc$^{-1}$, $\Omega_M=0.3$ and $\Omega_\Lambda=0.7$, and proto-Solar abundances (Lodders \& Palme \cite{Lod09}). The assumed statistics are C-statistics (Cash \cite{Cash79}). The quoted errors refer to 68.3\% confidence level ($\Delta C = 1$ for one parameter of interest) unless otherwise stated.

%__________________________________________________________________

\section{The X-ray data}
\label{Xdata}

%-----------------------------------
\begin{figure*}
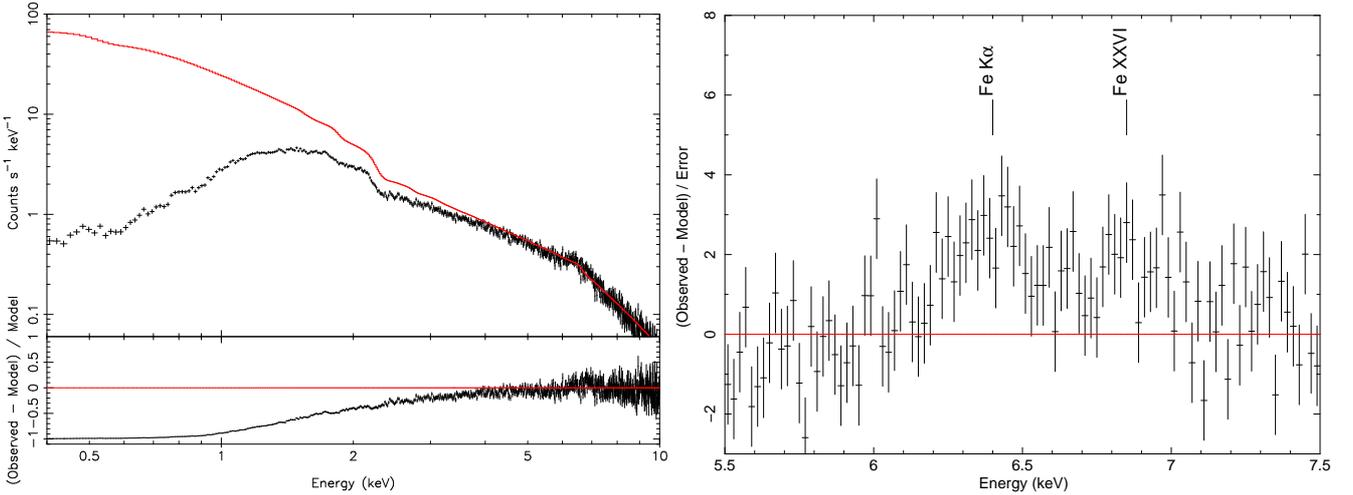

  \centering
  \hbox{
    \includegraphics[width=6.5cm,angle=-90]{17650fg1a.ps}
    \includegraphics[width=6.5cm,angle=-90]{17650fg1b.ps}
  }
  \caption{{\it Left panel:} EPIC-pn spectrum of NGC 7314 with residuals. The solid line represents the best-fit model excluding the neutral and ionized absorption and the Fe emission lines.  {\it Right panel:} Detail of the EPIC-pn spectrum residuals in the \FeKa{} region.}
  \label{pnspec}
\end{figure*}
%-----------------------------------

For this work we used X-ray data available in the archives of past observations of NGC 7314 made with \ASCA{}, \suz{}, and \XMM{} (see Table~\ref{obslog}).

The source was observed twice by \ASCA{} in 1994 and 1996 for $\sim$50~ks and $\sim$43~ks, respectively. We retrieved the SIS and GIS spectra from the Tartarus database, which were already binned so that there were at least 20 counts per channel.

NGC 7314 was also observed by \XMM{} in 2001 (ObsID 0111790101) for $\sim$44~ks. After removing flaring background periods the net exposure time was $\sim$30~ks. We extracted the EPIC-pn, which was operated in Small Window Mode, using the {\tt xmmselect} task under SAS v10.0.0. The source extraction region was 41 arcsec radius to prevent the extracting area to fall outside the data window. The background spectra was extracted using the same region size as the source, positioned in an area empty of X-ray sources. The spectrum was rebinned using the tool PHARBN (M. Guainazzi, private communication) so that the number of bins per resolution element of the pn spectrum was 3, and the minimum number of counts per channel was 20. The RGS spectra were extracted using the {\tt rgsproc} task with the standard settings for the size of the source and background regions.

The source was observed in 2007 by \suz{} (ObsID 702015010) for $\sim$108~ks. The \suz{} data were reduced following the \suz{} Data Reduction Guide (also known as the {\sl ABC} guide). We reprocessed all the data files using the standard screening settings. The net exposure time of the XIS detectors was $\sim$84~ks, and the CTI data of the XIS instrument was reprocessed using the {\tt xispi} task so that the latest calibration files were used at the moment of the analysis. XIS data were selected in $3 \times 3$ and $5 \times 5$ edit-modes using grades 0, 2, 3, 4, and 6. The $5 \times 5$ data were converted to $3 \times 3$ and merged with the original $3 \times 3$ data in order to increase the signal-to-noise ratio. The spectrum of NGC 7314 and the background spectrum were extracted using circular region of 1 arcmin radius. The spectra from the front-illuminated detectors XIS0 and XIS3 were combined to create a single source spectrum (task {\tt addascaspec}). Response and ancillary files were created using the tasks {\tt xisrmfgen} and {\tt xissimarfgen}, respectively.

%----------------------------
\begin{table}
  \centering
  \caption[]{NGC 7314 observation log.}
  \label{obslog}
  \begin{tabular}{l c c c}
    \hline\hline
    Observatory & Obs ID  &  Date & Net exp. time    \\
           &     &  (dd-mm-yyyy) &  (ks)  \\
    \hline
    \ASCA{} & 72029000 & 20-11-1994 & 45 \\
    \ASCA{} & 74059000 & 18-05-1996 & 42 \\
    \XMM{} & 0111790101 & 02-05-2001 & 30 \\
    \suz{} & 702015010 & 25-04-2007 & 84 \\
    \hline
  \end{tabular}
\end{table}
%----------------------------

%_________________________________________________________________

\section{Analysis}
\label{analysis}

In this section we present the spectral analysis of the \XMM{}, \suz{}, and \ASCA{} datasets. All the spectra were analyzed using the {\tt SPEX}\footnote{\tt http://www.sron.nl/spex} fitting package (Kaastra et al. \cite{Kaa96}) version 2.03.00.

\subsection{\XMM{}~EPIC-pn}
\label{pn}

We fitted the EPIC-pn spectrum continuum in the $0.4-10$~keV range with a single power-law corrected for the Galactic foreground absorption ($N_{\rm H}= 1.45 \times 10^{20}$~cm$^{-2}$, Dickey \& Lockman \cite{DL90}), which was kept fixed throughout the fits. The spectrum below 2~keV was heavily absorbed, so we introduced an absorption component at the redshift of the source. The neutral absorption is modeled with the {\tt hot} model in {\tt SPEX}, which calculates the transmission of a plasma in collisional ionization equilibrium, fixing the temperature to 0.5~eV and leaving the column density as the only free parameter. This fit yielded a column density for the neutral absorber of $N_{\rm H} = (7.0 \pm 0.1) \times 10^{21}$~cm$^{-2}$, but the fit still was unacceptable ($C/\nu=2249/638$) with several residuals below 2~keV and at the position of the Fe emission around 6.4~keV. We introduced a cold reflection component ({\tt refl} model in {\tt SPEX}) coupling the photon index $\Gamma$ to that of the power-law, leaving only the normalization and the emissivity $q$ as the only free parameters. The rest of the parameters of this component were left fixed at the default values of {\tt SPEX}. This improves the fit in $\Delta C/\Delta \nu= 308/2$ and successfully reproduces the \FeKa{} profile at $E \sim 6.4$~keV. The contribution of the reflection to the continuum is $\sim$4\%.

In order to account for the residuals at softer energies, we added an ionized absorber intrinsic to the source, following the preliminary analysis of Branduardi-Raymont et al. (\cite{BR02}). The ionized absorber is described by the {\tt xabs} model in {\tt SPEX}, which calculates the transmission of a slab of material where all ionic column densities are linked through an ionization balance model. We assume the default ionization balance of {\tt SPEX} and the default root mean square velocity $\sigma_{\rm v}$ of 100~km~s$^{-1}$. A single ionized absorber significantly improved the fit in $\Delta C/\Delta \nu= 749/2$. The reduced $C$ was, however, still too high and several residuals were still observed at soft X-rays. A second ionized absorber component was added further improving the fit by $\Delta C/\Delta \nu= 295/2$. Although visually good, some residuals remained below 1~keV so we tried to see if a lower ionization absorber was present by adding a third component, which rendered a further improvement of $\Delta C/\Delta \nu= 103/2$.

Hence, the spectrum of NGC 7314 reveals both intrinsic neutral and ionized absorbers along our line of sight. The neutral absorber has a column density of $N_{\rm H}= (2.9^{+0.3}_{-0.1}) \times 10^{21}$~cm$^{-2}$. The three ionized absorber components account for gas with three distinct ionization phases characterized by $\log \xi = 2.6 \pm 0.1$, $\log \xi = 1.2 \pm 0.1$, and $\log \xi = -1.3 \pm 0.3$, respectively. The ionization parameter $\xi$ is defined as $\xi = L/nR^2$, where $L$ is the luminosity of the ionizing continuum in the $1-1000$~Ryd range, $n$ is the density of the gas, and $R$ is the distance of the gas to the ionizing source. The column densities of the gas in these phases are $N_{\rm H} = (8.7^{+2.3}_{-0.8}) \times 10^{22}$~cm$^{-2}$, $N_{\rm H} = (3.5 \pm 0.3) \times 10^{21}$~cm$^{-2}$, and $N_{\rm H} = (7.2^{+0.1}_{-0.7}) \times 10^{22}$~cm$^{-2}$, respectively. The analysis of the EPIC-pn spectrum of Tombesi et al. (\cite{Tom10}) rendered a slighly flatter continuum ($\Gamma = 1.97 \pm 0.04$) and cold absorption with $N_{\rm H} = (1.8 \pm 0.4) \times 10^{22}$~cm$^{-2}$. We note, however, that this analysis was carried out only in the $4 - 10$~keV energy band, and the soft X-ray spectrum, where the ionized absorption is most significant, was not taken into account.

\begin{figure}
  \centering
  \includegraphics[width=6.5cm,angle=-90]{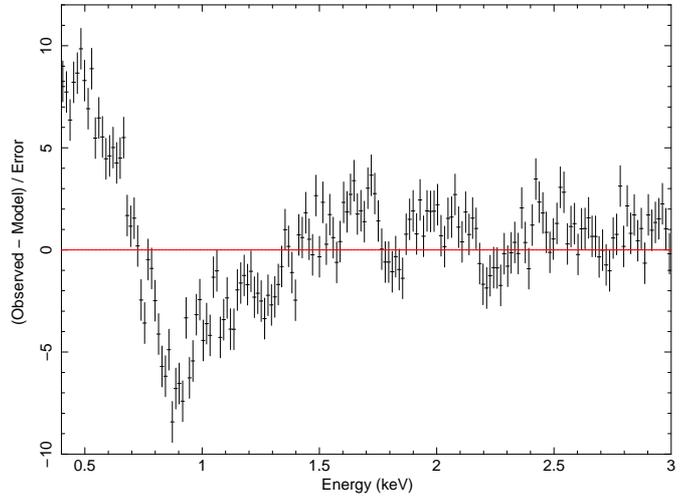}
  \caption{Detail of the EPIC-pn residuals in the soft X-ray band after fitting the continuum plus intrinsic neutral absorber.}
  \label{pnsoftres}
\end{figure}

An excess around $\sim$6.9~keV was modeled adding a Gaussian line at this position, being the free parameters the line energy, the FWHM, and the normalization of the line. We identify this line as emission from Fe\,{\sc xxvi}. The addition of this line improved the fit in $\Delta C/\Delta \nu= 26/3$, which translates into a significance probability of $\sim$99.6\%. Both Fe K$\alpha$ and Fe\,{\sc xxvi} lines are unresolved. The centroid of the Fe K$\alpha$ line is $E=6.40 \pm 0.03$~keV, while for the Fe\,{\sc xxvi} line is $E=6.91 \pm 0.04$~keV (see Fig.~\ref{pnspec}, right panel). The lines have a FWHM of $0.40 \pm 0.10$~keV and $0.21 \pm 0.11$~keV, respectively. These widths are consistent within $1\sigma$, but the slightly broader \FeKa{} line suggests that likely originates by reflection in the disk whereas the \FeXXVI{} line probably originates at a more distant location. In the left panel of Fig.~\ref{pnspec} we show the EPIC-pn spectrum of NGC 7314 with the best-fit model excluding the neutral and ionized absorbers and the Fe emission lines. A detailed view of the Fe region is shown in the right panel of Fig.~\ref{pnspec}. A detail of the EPIC-pn spectrum in the $0.4-3$~keV range after fitting the continuum and the intrinsic neutral absorber is shown in Fig.~\ref{pnsoftres}. The formal final best fit is therefore $C/\nu=768/627$, and the results of the fit are summarized in Table~\ref{fitstable}.

%-----------------------------------
%-----------------------------------

\subsection{\XMM{}~RGS}
\label{rgs}

A more detailed view of these absorbers could, in principle, be obtained with the XMM-\emph{Newton} RGS. Unfortunately, the high absorption in soft X-rays, where the effective area of RGS peaks, severely affects the statistics of the spectrum leaving very few distinguishable features.

A preliminary fit to the RGS spectrum (both RGS1 and RGS2 were fitted together) with a single power-law plus two neutral absorbers, one fixed to the Galactic value and one intrinsic to the source yields a visually unacceptable fit with $C/\nu=256/229$. The addition of an ionized absorber with $\log \xi \sim 1$ at the redshift of the source improves the fit by $\Delta C/\Delta \nu= 24/2$ (significance probability of $\sim$99.9\%). A second ionized absorber with $\log \xi \sim 2.4$ further improves the fit by $\Delta C/\Delta \nu= 20/2$ (significance probability of $\sim$99.8\%). The addition of a low-ionization phase in the ionized absorber, similar to the one that we found in the pn spectrum no longer improves the fit. Likewise, the presence of a reflection component in the continuum is not statistically significant because most features of this component fall outside the RGS waveband.

\begin{figure}
  \centering
  \includegraphics[width=6.5cm,angle=-90]{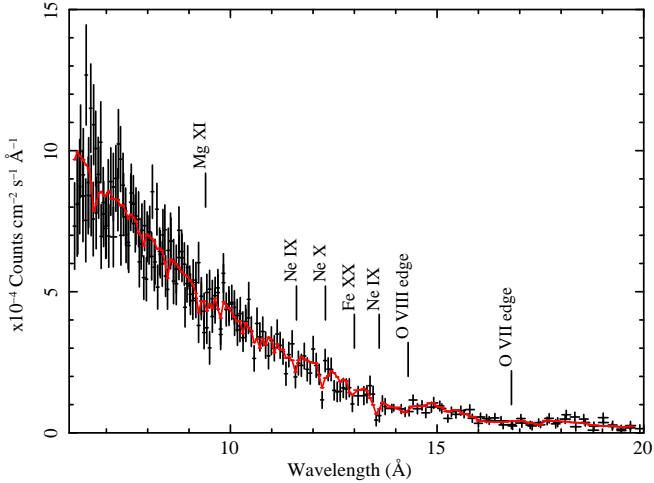}
  \caption{RGS spectrum of NGC 7314. The solid line represents the best-fit model, and some of the most significant features are labeled.}
  \label{rgsspec}
\end{figure}

Therefore, the final RGS spectrum fit (see Table~\ref{fitstable}~and Fig.~\ref{rgsspec}) shows high intrinsic neutral absorption ($N_{\rm H}= 6.0 \pm 0.4 \times 10^{21}$~cm$^{-2}$), and the presence of, at least, two ionized absorbers ($\log \xi = 2.4 \pm 0.2$ and $\log \xi = 1.0 \pm 0.2$, respectively) with column densities of $N_{\rm H} = (7.5 \pm 2.6) \times 10^{21}$~cm$^{-2}$ and $N_{\rm H} = (2.7 \pm 0.8) \times 10^{21}$~cm$^{-2}$.

The low-ionization phase seen in the EPIC-pn spectrum is statistically undetected in RGS due to the low number of counts at longer wavelengths. This explains the higher $N_{\rm H}$ of the neutral absorber measured in this spectrum with respect to the pn one, as it can account for the blend of neutral and the low-ionization gas. However, this is still consistent with the complex absorption scenario found in the pn spectrum.

%-----------------------------------
%-----------------------------------

%-----------------------------------
\begin{figure*}
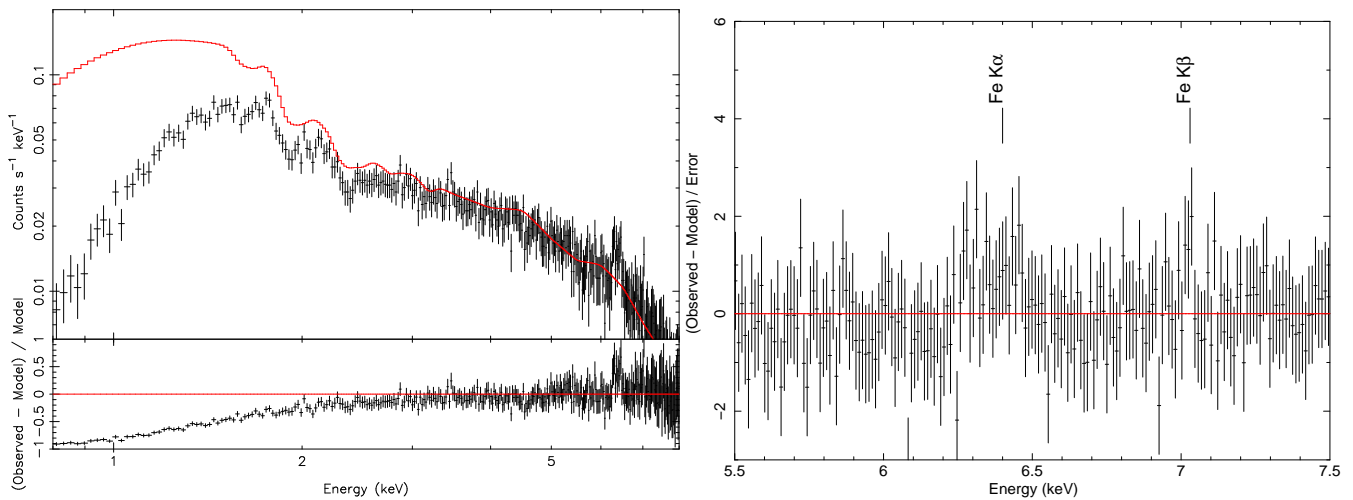

  \centering
  \hbox{
    \includegraphics[width=6.5cm,angle=-90]{17650fg4a.ps}
    \includegraphics[width=6.5cm,angle=-90]{17650fg4b.ps}
  }
  \caption{{\it Left panel:} \suz{} XIS spectrum with residuals of NGC 7314. The solid line represents the power-law best-fit model excluding intrinsic absorption and Fe emission lines. {\it Right panel:} Detail of the \suz{} XIS spectrum residuals in the \FeKa{} region.}
  \label{xisspec}
\end{figure*}
%-----------------------------------

%-----------------------------------
\begin{figure*}
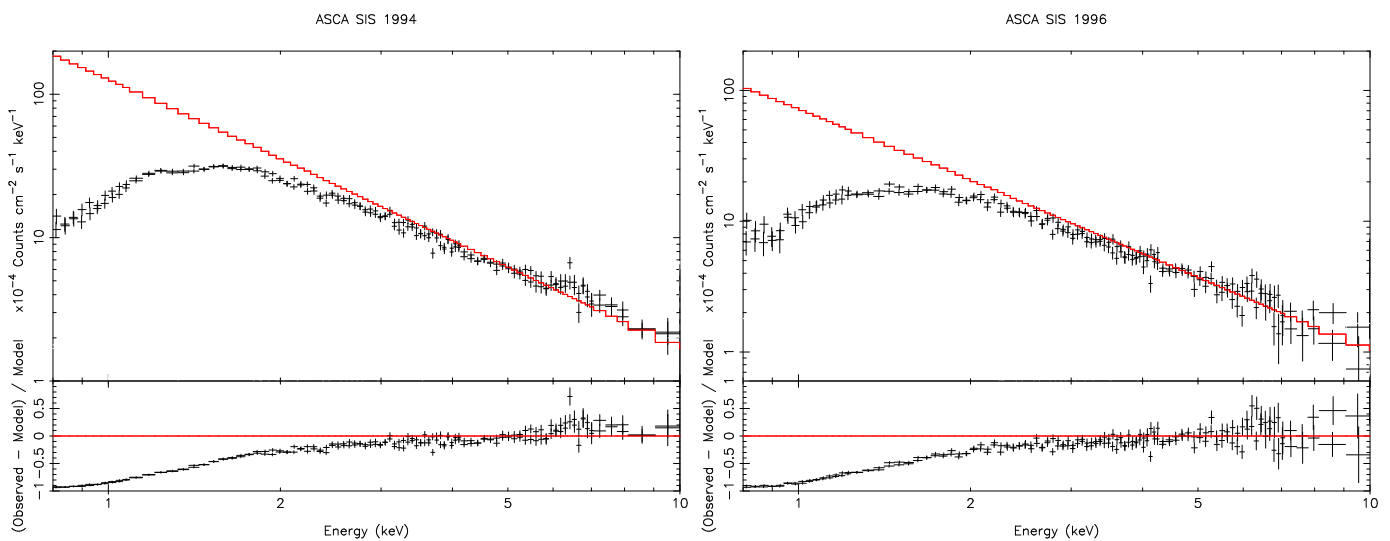

  \centering
  \hbox{
    \includegraphics[width=7cm,angle=-90]{17650fg5a.ps}
    \includegraphics[width=7cm,angle=-90]{17650fg5b.ps}
  }
  \caption{\ASCA{} SIS spectra and residuals of NGC 7314 in the 1994 ({\it left panel}) and 1996 ({\it right panel}) observations. The solid line represents the power-law best-fit model to both SIS and GIS spectra excluding intrinsic absorption and the \FeKa{} emission line. For clarity only the SIS0 and SIS1 spectra of each observation are plotted.}
  \label{ascaspec}
\end{figure*}
%-----------------------------------

\subsection{\suz{}~XIS}
\label{xis}

Similarly to the EPIC-pn spectrum, the continuum of the \suz{} XIS spectrum was first modeled in the $0.8-10$~keV range using a single power-law corrected for local Galactic absorption. Since our aim for this paper was to study the low energy absorption components, we did not include the PIN data in the analysis (see De Marco et al., in preparation, for the broad band spectral fits and discussion).

As expected, large residuals were found at energies $<$2~keV due to the large absorption ($C/\nu=2391/655$). After the addition of neutral absorption intrinsic to NGC 7314 the fit improved significantly ($C/\nu=778/654$). Interestingly, the intrinsic ionized absorber that we significantly found in the \XMM{} spectra seems to be absent or marginally present in this \suz{} dataset. Our attempts to fit a single {\tt xabs} component using any of the best-fit values in the EPIC-pn spectrum led to a drop in the column densities to zero for the high- and medium-ionization components. Only the low-ionization component could be fitted with $N_{\rm H} = 5.9 \pm 1.4 \times 10^{21}$~cm$^{-2}$ and $\log \xi = 0.0 \pm 0.5$, although it only marginally improved the overall fit ($\Delta C/\Delta \nu= 8/2$, for a significance probability of 95.4\%). Moreover, the column density of the neutral absorber is almost three times larger in the XIS spectrum than in the pn one ($N_{\rm H} = (8.9 \pm 0.2) \times 10^{21}$~cm$^{-2}$), and continuum slope is significantly flatter ($\Gamma = 1.82 \pm 0.02$). This will be further discussed in Sect.~\ref{varabs}.

An unresolved \FeKa{} emission line is also significantly detected in the \suz{} XIS spectrum, and can be successfully accounted for after including a reflection component to the continuum in the same manner as described for the EPIC-pn spectrum. The fit improvement is $\Delta C/\Delta \nu= 46/2$. We identify the narrow excess above 7~keV as emission from \FeKb{}, which is detected at the $3\sigma$ level, and modeled with a delta line ($\Delta C/\Delta \nu= 17/2$, significance probability of $\sim$99.7\%). However, the flux of the \FeKb{} line is $\sim4$ times larger than expected, which may indicate some contamination. This aspect will be discussed in Sect.~\ref{interp}. The XIS spectrum fit is summarized in Table~\ref{fitstable}, and is shown in Fig.~\ref{xisspec}. The \suz{} spectrum of NGC 7314 was also analyzed by Fukazawa et al. (\cite{Fuk11}) as part of a study of the Fe K lines in a sample of 88 Seyfert galaxies. They report an unresolved \FeKa{} line at $E = 6.39 \pm 0.01$~keV with ${\rm FWHM} = 0.11 \pm 0.05$~keV, and a possible \FeKb{} line at $E \sim 7$~keV with ${\rm FWHM} = 0.07 \pm 0.02$~keV, fully in agreement with our results. Since they focus in the $5-9$~keV range, they do not provide any discussion on the properties of the absorber in the source.

%-----------------------------
\begin{table*}
  \renewcommand{\arraystretch}{1.6}
  \centering
  \caption[]{Best-fit summary to the different NGC 7314 spectra.}
  \label{fitstable}
  \begin{tabular}{l l c c c c c}
    \hline\hline
    Component  &  Free parameter  &       EPIC-pn  &      RGS     &           XIS    &           SIS/GIS 94    &    SIS/GIS 96 \\
    \hline
    Power-law & ${\rm Flux}$ ($2-10$~keV)\tablefootmark{a} & $3.91 \pm 0.04$ & $5.9 \pm 1.0$ & $0.89 \pm 0.02$ & $3.71 \pm 0.05$ & $2.3 \pm 0.05$ \\
              & $\Gamma$              & $2.14 \pm 0.01$ & $1.67 \pm 0.19$ & $1.82 \pm 0.02$   & $2.01 \pm 0.01$ & $1.96 \pm 0.01$ \\
    \hline
    Reflection & ${\rm Flux}$ ($2-10$~keV)\tablefootmark{a} & $0.16 \pm 0.03$ & $\dots$          & $0.014 \pm 0.001$  & $0.16 \pm 0.02$  & $0.104 \pm 0.004$  \\
               & $\Gamma$   & $2.14$\tablefootmark{f} & $\dots$ & $1.82$\tablefootmark{f} & $2.01$\tablefootmark{f} & $1.96$\tablefootmark{f} \\
               & $q$\tablefootmark{b} & $-2.2 \pm 0.1$ & $\dots$ & $-1.7^{+0.4}_{-0.2}$ & $-1.9 \pm 0.2$ & $-2.1 \pm 0.2$ \\
    \hline
    Neutral abs. & $N_{\rm H}$\tablefootmark{c} & $2.9^{+0.3}_{-0.1}$ & $6.0 \pm 0.4$ & $8.9 \pm 0.2$ & $9.4 \pm 0.1$ & $9.5 \pm 0.2$ \\
    \hline
    Ionized abs. & $N_{\rm H}$\tablefootmark{c} & $7.2^{+0.1}_{-0.7}$ & $\dots$ & $\dots$ & $\dots$ & $\dots$ \\
                 & $\log \xi$\tablefootmark{d} & $-1.3 \pm 0.3$ & $\dots$ & $\dots$ & $\dots$ & $\dots$ \\
    \hline
    Ionized abs. & $N_{\rm H}$\tablefootmark{c} & $3.5 \pm 0.3$ & $2.7 \pm 0.8$ & $\dots$ & $\dots$ & $\dots$ \\
                 & $\log \xi$\tablefootmark{d} & $1.2 \pm 0.1$ & $1.0 \pm 0.2$ & $\dots$ & $\dots$ & $\dots$ \\
    \hline
    Ionized abs. & $N_{\rm H}$\tablefootmark{c} & $8.7^{+2.3}_{-0.8}$ & $7.5 \pm 2.6$ & $\dots$ & $\dots$ & $\dots$ \\
                 & $\log \xi$\tablefootmark{d} & $2.6 \pm 0.1$ & $2.4 \pm 0.2$ & $\dots$ & $\dots$ & $\dots$ \\
    \hline
    \FeKa{}      & ${\rm Flux}$ ($2-10$~keV)\tablefootmark{a} & $0.019 \pm 0.003$ & $\dots$ & $0.006 \pm 0.001$ & $0.037^{+0.038}_{-0.010}$ & $0.035 \pm 0.007$ \\
                 & $E$\tablefootmark{e}    & $6.40 \pm 0.03$ & $\dots$ & $6.41 \pm 0.02$ & $6.47^{+0.06}_{-0.11}$ & $6.43 \pm 0.07$ \\
                 & ${\rm FWHM}$\tablefootmark{e} & $0.40 \pm 0.10$ & $\dots$ & $0.11 \pm 0.05$ & $0.82^{+1.15}_{-0.35}$ & $0.85 \pm 0.20$ \\
    \hline
    \FeXXVI{}    & ${\rm Flux}$ ($2-10$~keV)\tablefootmark{a} & $0.009 \pm 0.002$ & $\dots$ & $\dots$ & $\dots$ & $\dots$ \\
                 & $E$\tablefootmark{e}    & $6.91 \pm 0.04$ & $\dots$ & $\dots$ & $\dots$ & $\dots$ \\
                 & ${\rm FWHM}$\tablefootmark{e} & $0.21 \pm 0.11$ & $\dots$ & $\dots$ & $\dots$ & $\dots$ \\
    \hline
    \FeKb{}      & ${\rm Flux}$ ($2-10$~keV)\tablefootmark{a} & $\dots$ & $\dots$ & $0.0026 \pm 0.0006$ & $\dots$ & $\dots$ \\
                 & $E$\tablefootmark{e}    & $\dots$ & $\dots$ & $7.05 \pm 0.02$ & $\dots$ & $\dots$ \\
                 & ${\rm FWHM}$\tablefootmark{e} & $\dots$ & $\dots$ & $< 0.1$ & $\dots$ & $\dots$ \\
    \hline
  \end{tabular}
  \tablefoot{
    \tablefoottext{a}{In units of $10^{-11}$~erg~s$^{-1}$~cm$^{-2}$.}
    \tablefoottext{b}{Disk emissivity.}
    \tablefoottext{c}{Column density, in units of $10^{21}$~cm$^{-2}$.}
    \tablefoottext{d}{Ionization parameter, in units of erg~cm~s$^{-1}$.}
    \tablefoottext{e}{In units of keV.}
    \tablefoottext{f}{Parameter coupled to the $\Gamma$ of the power-law component.}
  }
\end{table*}
%-----------------------------

\subsection{\ASCA{}~SIS/GIS}
\label{asca}

We also analyzed the spectra of the solid-state imaging spectrometer (SIS) and gas imaging spectrometer (GIS) on board \ASCA{} for the 1994 and 1996 observations. We performed a broad-band fit of the SIS spectra in the $0.8-10$~keV range, and in the $1.4-10$~keV range for the GIS spectra. In both 1994 and 1996 observations the SIS0, SIS1, GIS2 and GIS3 spectra were fitted together.

The 1994 combined SIS/GIS spectra were modeled using a single power-law corrected by local absorption and intrinsic neutral absorption ($C/\nu=308/182$). Like in the case of the \suz{} XIS spectrum, the addition of one or more ionized absorbers to the model do not statistically improve the fit, the column densities dropping to zero when performing the fit. We note that an ionized absorber with column densities as large as $\sim 10^{22}$~cm$^{-2}$, like the one detected in the \XMM{} spectra, would also be easily detected by \ASCA{} if it was present (e.g. George et al. \cite{Geo98}). Therefore, the WA is either absent or it has a much lower $N_{\rm H}$. The addition of a reflection component gives the formal final best-fit of $C/\nu=590/425$. At higher energies only the \FeKa{} emission line is present. The 1996 \ASCA{} SIS/GIS spectra were modeled in a similar way for a formal best-fit of $C/\nu=628/435$. The \ASCA{} SIS/GIS best-fit parameters of both observations are reported in Table~\ref{fitstable}. The SIS spectra of NGC 7314 for the 1994 and 1996 observations are shown in Fig.~\ref{ascaspec}.

The best-fit parameters for the continuum and absorption models are consistent within the error bars between the 1994 and 1996 observations, and likewise for the \FeKa{} parameters. The \FeKa{} line shows a broad component that was first reported in Yaqoob et al. (\cite{Yaq96}). In particular, in the 1996 \ASCA{} observation, the Fe K region looks noisy (see Fig.~\ref{ascaspec}, right panel). The ${\rm FWHM}$ of the \FeKa{} line obtained from the fits of the SIS and GIS independently gives $0.36 \pm 0.19$ and $0.98 \pm 0.22$~keV, respectively, while the simultaneous fit of all the spectra gives $0.85 \pm 0.20$ (see Table~\ref{fitstable}). This may indicate that the \FeKa{} profile likely consists of a narrow core plus broad wings similarly as in the Yaqoob et al. (\cite{Yaq96}) analysis of the 1994 spectrum. An alternative possibility where the \FeKa{} profile could be due to a blend of unresolved narrow component cannot be ruled out, but it is difficult to statistically discriminate between both possibilities with the current resolution. The differences in the best-fit parameters between the \ASCA{}, \suz{} and \XMM{} observations are further discussed in the next section.

%_________________________________________________________________

\section{Discussion}
\label{discussion}

\subsection{Spectral variability in NGC 7314}
\label{specvar}

NGC 7314 is known for showing very rapid short-term variability on timescales of hundreds of seconds (see e.g. Yaqoob et al. \cite{Yaq96}, Zou \& Xue \cite{ZX02}). In addition to this, the source experiences long-term variability (of the order of years) between the different observations. In Table~\ref{fitstable} we report the flux of NGC 7314 in the $2-10$~keV band as obtained from the different spectral fits. It can be seen that between the \ASCA{} observations of 1994 and 1996 there is a drop in flux of $\sim$40\%, followed by a recovery of the same amount at the time of the \XMM{} observation 5 years later. However, the most significant variation in flux occurs between the \XMM{} and \suz{} observations, with a drop in flux of a factor of 4. These flux variations are correlated with changes in the best-fit photon index $\Gamma$, so that the spectrum steepens as the flux increases.

Interestingly, the decrease in flux by a factor of $\sim$4 between the \XMM{} and \suz{} observations corresponds also to an increase in the neutral column density $N_{\rm H}$ by a factor of $\sim$3 between these observations. This fits well in a scenario where the variability in the primary radiation is the main driver of the observed absorber variations. However, in the \ASCA{} data the flux is of the same order as in the \XMM{} observation, while the measured neutral absorber is of the order of that measured in the \suz{} observation with no significant detection of an ionized absorber. This situation seems incompatible with the above scenario, thus indicating that other mechanism besides the flux variability might be responsible for the changes in the absorbers.

The timing properties of NGC 7314 at both short- and long-term timescales will be further discussed in a forthcoming paper (De Marco et al., in preparation).

\subsection{Variable absorption in NGC 7314}
\label{varabs}

%-----------------------------------
\begin{figure*}
  \centering
  \hbox{
  \includegraphics[width=6.5cm,angle=-90]{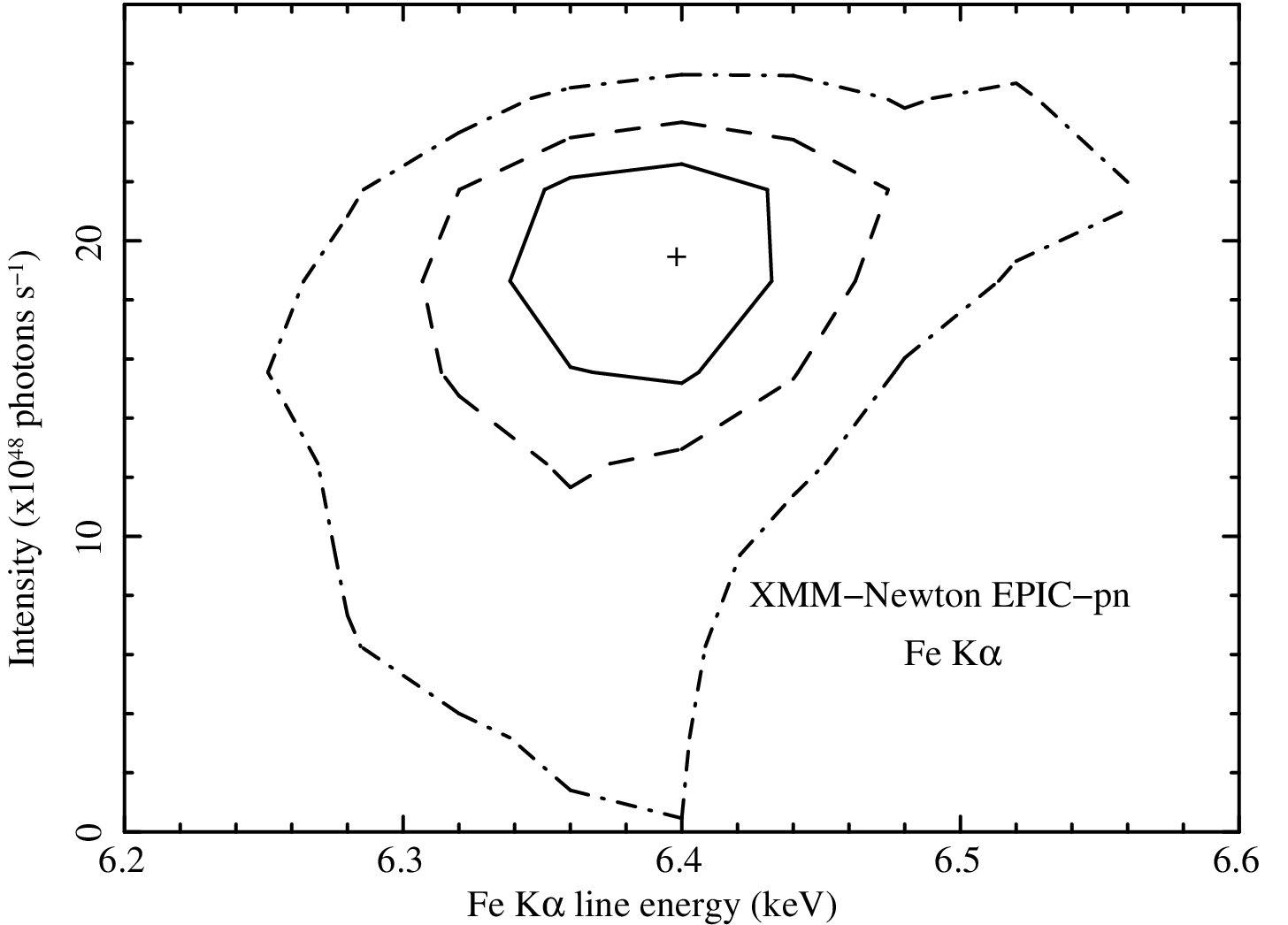}
  \includegraphics[width=6.5cm,angle=-90]{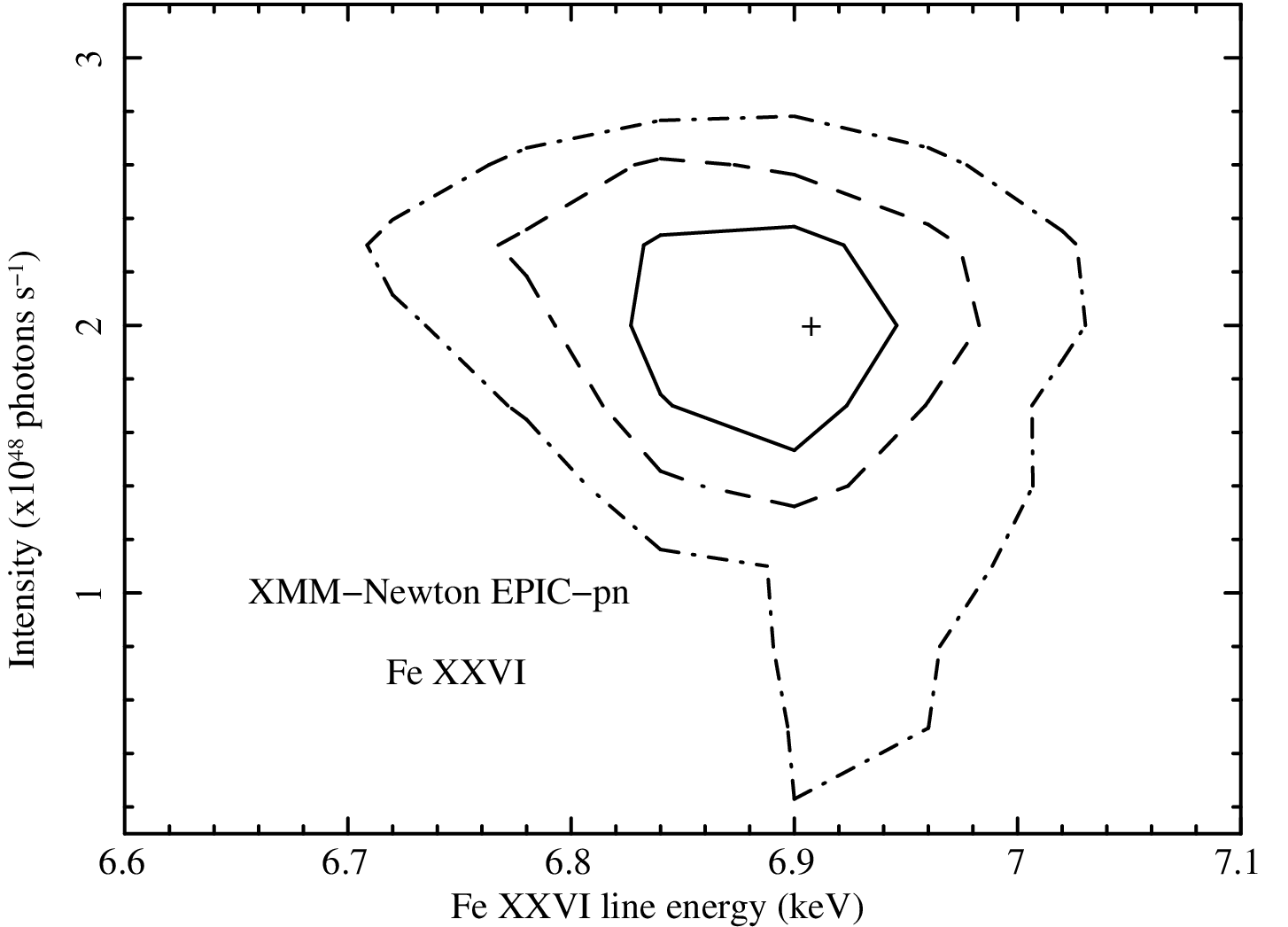}
  }
  \caption{The $1\sigma$ (solid line), $2\sigma$ (dashed line), and $3\sigma$ (dot-dashed line) confidence contours of the line energy versus the line intensity for \FeKa{} ({\it left panel}) and \FeKb{} ({\it right panel}) in the \XMM{} pn spectrum.}
  \label{fecontourspn}
\end{figure*}
%-----------------------------------
%-----------------------------------
\begin{figure*}
  \centering
  \hbox{
  \includegraphics[width=6.5cm,angle=-90]{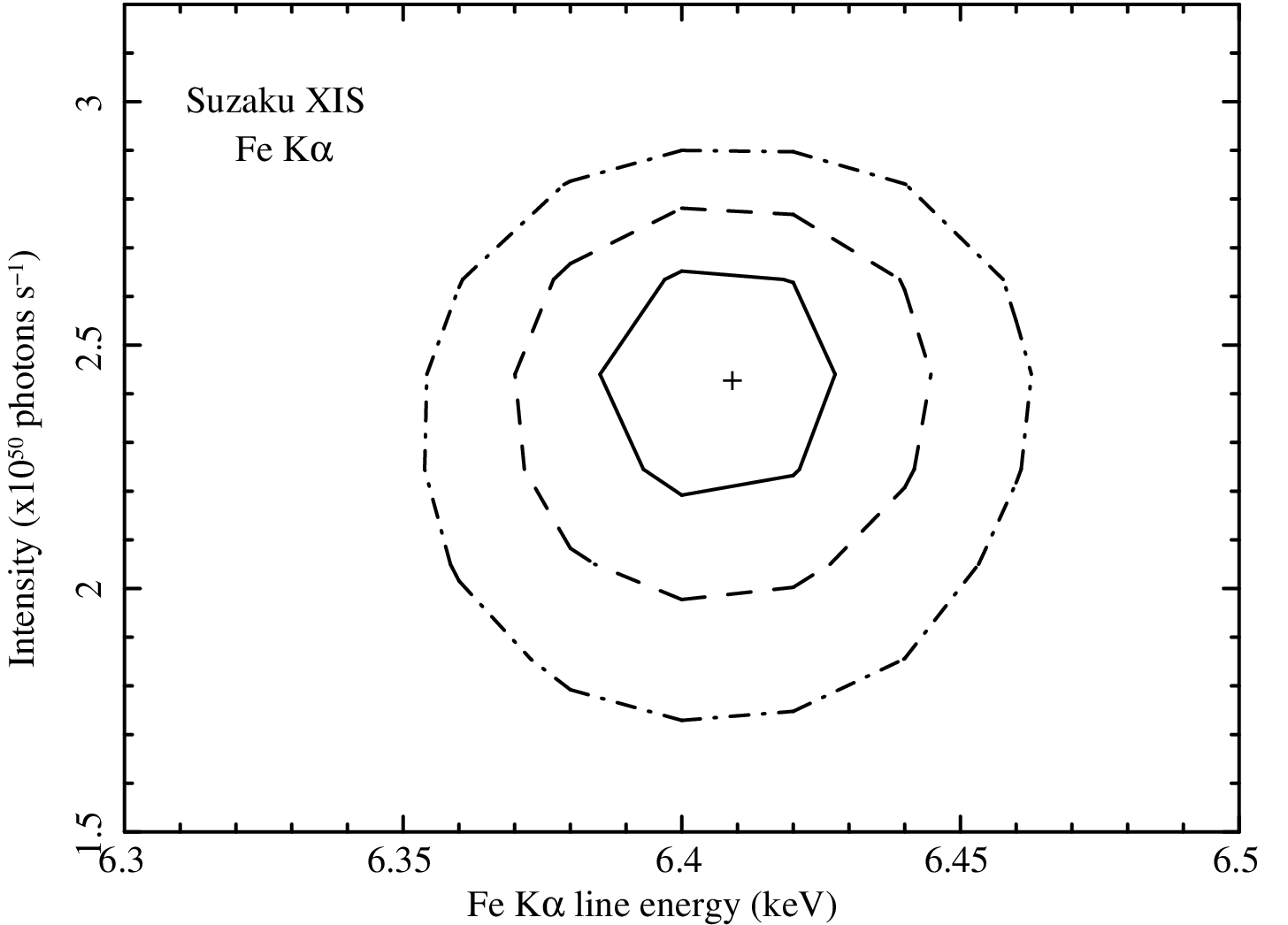}
  \includegraphics[width=6.5cm,angle=-90]{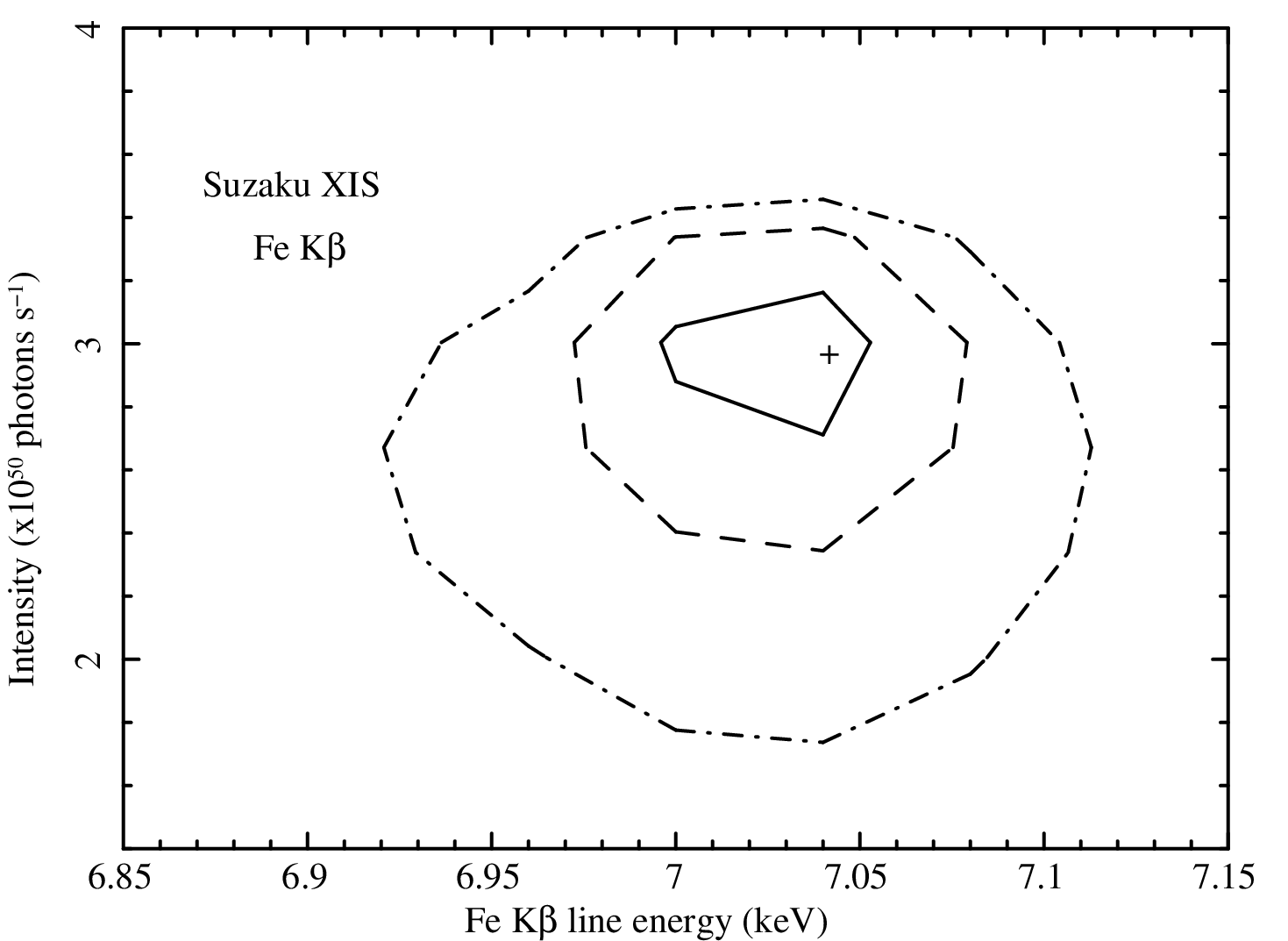}
  }
  \caption{The $1\sigma$ (solid line), $2\sigma$ (dashed line), and $3\sigma$ (dot-dashed line) confidence contours of the line energy versus the line intensity for \FeKa{} ({\it left panel}) and \FeKb{} ({\it right panel}) in the \suz{} XIS spectrum.}
  \label{fecontours}
\end{figure*}
%-----------------------------------

From the analysis of the different observations, we see that the bulk of the absorption in the \XMM{} observation is caused by an ionized absorber which is not present or is statistically undetectable in the \ASCA{} and, particularly, the deeper \suz{} observation. This poses the question of whether the measured absorption in NGC 7314 has genuinely varied or the changes in the absorber are due to an unresolved low-ionization component.

Indeed, from the \suz{} data analysis (see Sect.~\ref{xis}) we see that the presence of a highly ionized absorber can be statistically ruled out. For a {\tt xabs} component with $\log \xi$ fixed to 2.6, the value we obtained for the EPIC-pn spectrum, it gives an upper limit for the column density of this component of $N_{\rm H} < 2 \times 10^{21}$~cm$^{-2}$, well below the $N_{\rm H}$ measured in EPIC-pn. On the other hand, a mildly ionized warm absorber with $\log \xi \simeq 0$ was marginally detected. In this case, the column density of the WA is $N_H = (5.2 \pm 2.2) \times 10^{21}$~cm$^{-2}$, while that of the neutral absorber is $N_H = (5.7 \pm 1.2) \times 10^{21}$~cm$^{-2}$. This WA could be an unresolved mixture of the low- and medium-ionization gas detected in the EPIC-pn spectrum, but its presence cannot be statistically confirmed (the fit improves only $\Delta C/\Delta \nu= 8/2$, while the combined low and medium WA in the EPIC-pn spectrum provides an overall improvement of $\Delta C/\Delta \nu= 398/4$). 

A further test in which we assumed the WA in the \suz{} XIS spectrum to be composed only by the low-ionization gas detected in pn ($\log \xi$ fixed to $-1.3$), yielded a WA column density of $N_H = (7.1 \pm 3.2) \times 10^{21}$~cm$^{-2}$. Interestingly, in this case the neutral absorption we measure $N_H = (4.0 \pm 1.9) \times 10^{21}$~cm$^{-2}$ is fully consistent with that of \XMM{} ($N_H = (3.2 \pm 0.3) \times 10^{21}$~cm$^{-2}$), albeit with much larger error bars. The improvement of the fit in this case is also marginal $\Delta C/\Delta \nu= 5/1$, and it does not explain the large difference in the total absorbing column density between the \suz{} and \XMM{} observations due to the lack of high-ionization absorbing gas. This result is therefore likely to be consistent with a variation in the absorption properties of NGC 7314.

\subsection{A test for the unified model}
\label{interp}

%-----------------------------------
\begin{figure*}
  \centering
  \hbox{
    \includegraphics[width=5.5cm,angle=-90]{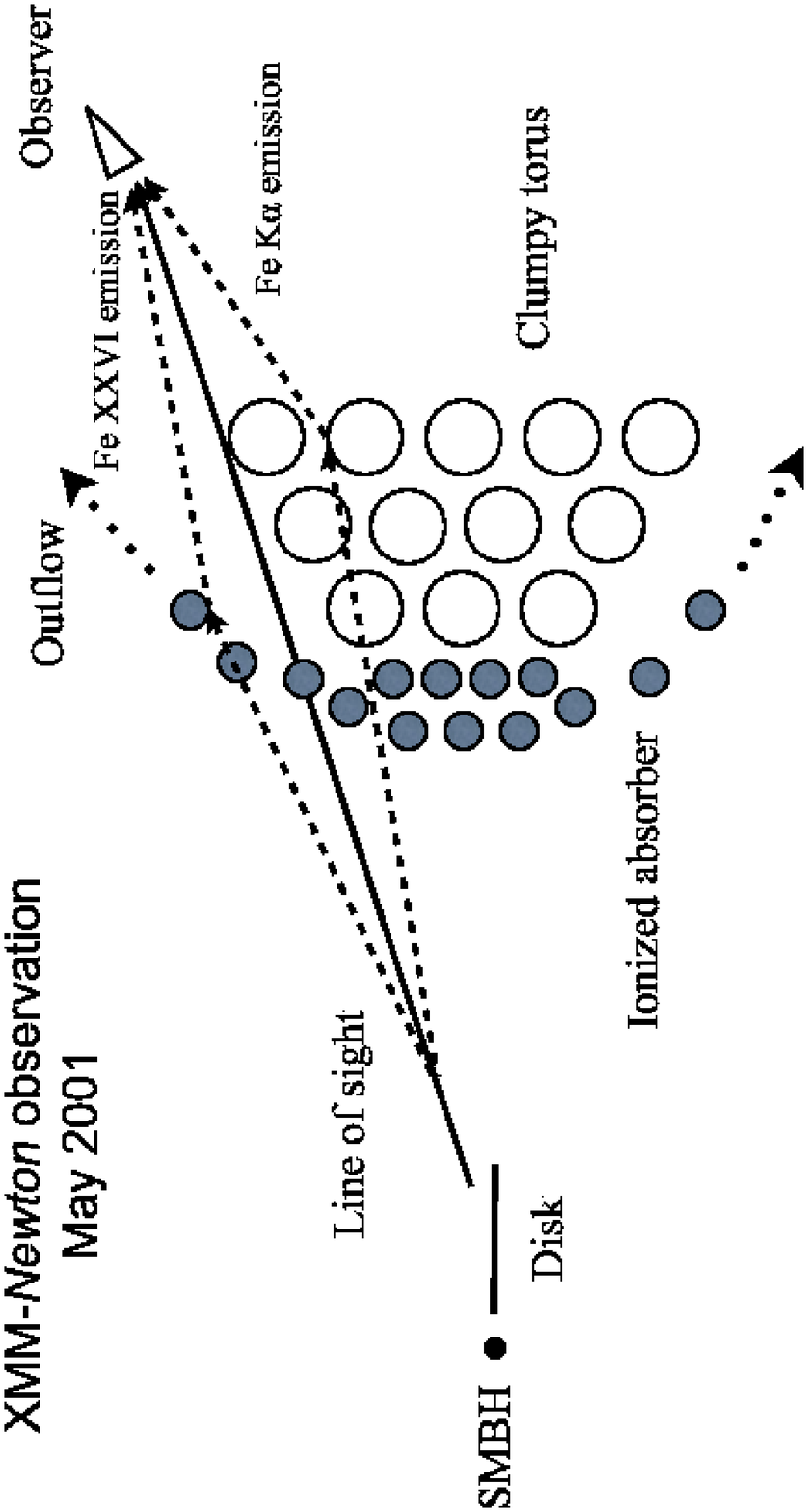}
    \includegraphics[width=5.5cm,angle=-90]{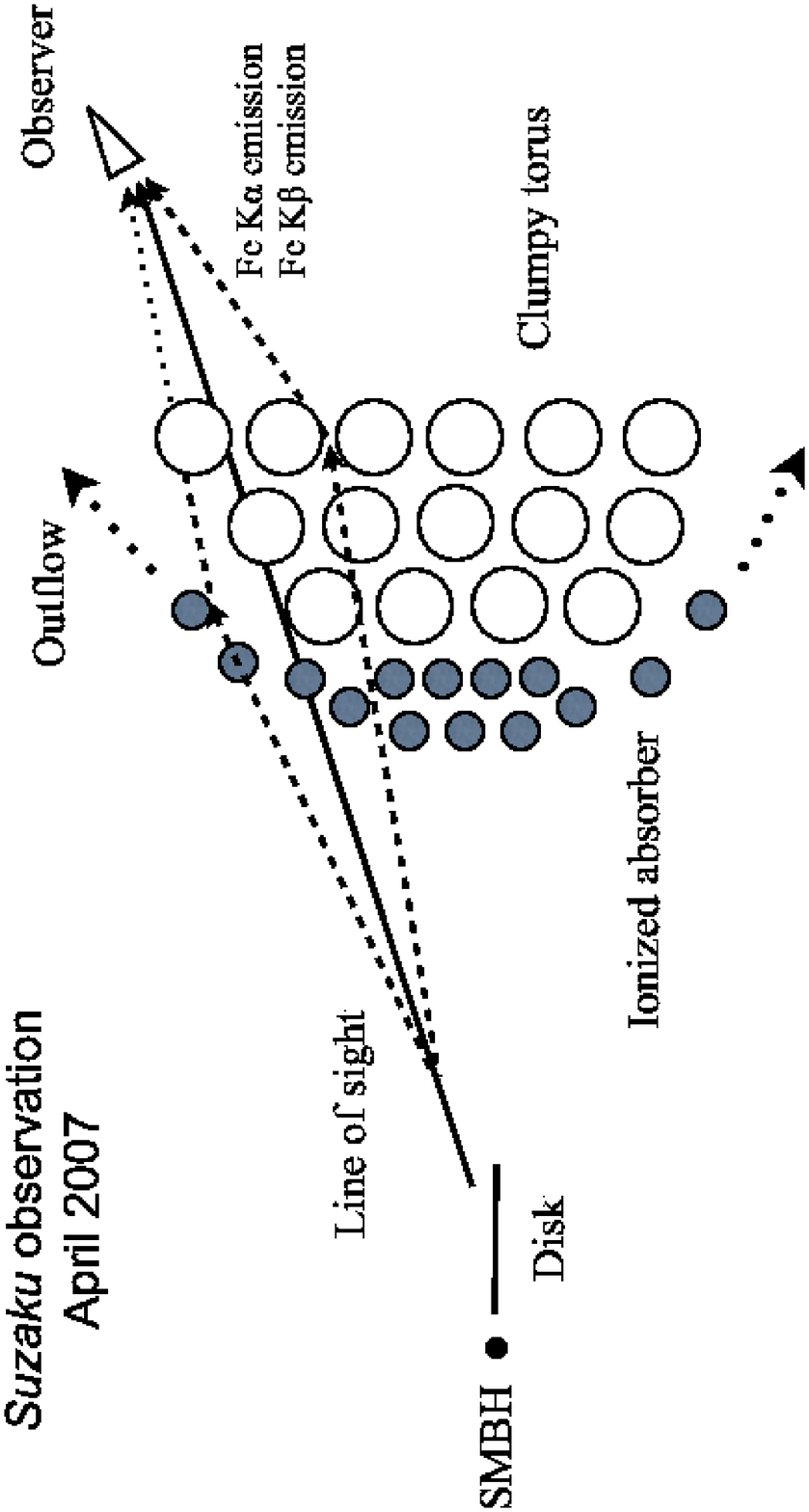}
  }
  \caption{Sketch of the inner regions of the AGN in NGC 7314. For clarity, the different parts are not to scale. {\it Left panel:} Situation at the moment of the \XMM{} observation. The line of sight grazes the clumpy torus, allowing the observer to detect the ionized absorber, which is located within the torus. Emission lines from \FeKa{} and \FeXXVI{} can be detected. {\it Right panel:} Situation at the moment of the \suz{} observation. The line of sight grazes the clumpy torus. A cloud of neutral material intercepts our line of sight, thus obscuring the inner regions to the observer and making the detection of a WA much more difficult. Emission lines from \FeKa{} and \FeKb{} can be detected.}
  \label{sketch}
\end{figure*}
%-----------------------------------

In the previous sections we have shown that the absorption spectrum of NGC 7314 presents different properties over the different observations. The \ASCA{} and \suz{} spectra, taken almost 13 years apart, show great similarities, being their absorption properties (absorption caused by neutral gas along our line of sight) fully consistent with each other. On the other hand, the \XMM{} observation in 2001, in between the \ASCA{} and \suz{} observations, strongly differs from them, as there are significant evidence of ionized absorption in the soft X-ray spectrum.

Moreover, there are also noticeable differences in the Fe K region of NGC 7314. The presence of the \FeKa{} emission line is significant in all of the observations. The \FeKa{} line profile is slightly broader in the \ASCA{} observations with respect to the \XMM{} and \suz{} observations. Yaqoob et al. (\cite{Yaq96}) already suggested in their analysis of the \ASCA{} spectra that the broad wing could originate close to the central engine, while the narrow core was likely originating from reflection in more distant matter.

Near to the \FeKa{} line, at higher energies, there is another emission feature in the EPIC-pn spectrum, which we identify with \FeXXVI{} (see Sect.\ref{pn}, and Fig.~\ref{fecontourspn}), detected at $\sim 3\sigma$ confidence level. This line, of similar width to that of the \FeKa{} line, is not present in the \suz{} XIS spectrum, which instead shows evidence of a narrow \FeKb{} emission line, also detected at the $\sim 3\sigma$ level (see Fig.~\ref{fecontours}). The \FeXXVI{} line is statistically undetected in the \suz{} XIS spectrum but it still could be very weakly present.

In Fig.~\ref{fecontourspn} we show the $1\sigma$, $2\sigma$, and $3\sigma$  confidence contours for the line intensities of \FeKa{} and \FeXXVI{} versus the line energy measured in the EPIC-pn spectrum. Likewise in Fig.~\ref{fecontours} for the \FeKa{} and \FeKb{} lines in the \suz{} XIS spectrum. In the first case, the $3\sigma$ contour of the \FeKa{} line extends towards higher energies, while that of the \FeXXVI{} line extends towards lower energies, which may indicate that a \FeXXV{} line could be present at $\sim6.6$~keV, although it is statistically undetected. 

In the case of \suz{} the narrow core of \FeKa{} is well constrained, whereas the $3\sigma$ contour of the \FeKb{} line is slightly elongated towards lower energies, which may suggest a marginal presence of \FeXXVI{} which was statistically undetected in the spectral fits (see Sect.~\ref{xis}). However, the unusually higher flux measured for \FeKb{} may indicate that some contamination is in place. Furthermore, the detection of the \FeKb{} line means that the gas responsible for the emission of this line, and the \FeKa{} line, has to be less ionized than Fe\,{\sc xvii}, because this ion and higher ionized Fe ions do not have M-shell electrons (e.g. Yaqoob et al. \cite{Yaq07}). This in addition with the lack of a significant \FeXXVI{} line favors a scenario with more neutral matter along our line of sight compared to the previous \XMM{} observation. Moreover, the broad component of the \FeKa{} line that is clearly seen in the \ASCA{} observations is less significant in the \XMM{} observation (although the widths are consistent within $1-2 \sigma$), while it is statistically undetected in the \suz{} spectrum. An increase in the column density of neutral matter along our line of sight in the latter case would then make more difficult to detect the broad component, which likely originates in the inner regions of the AGN, closer to the central super-massive black hole.

The unified model of AGN interprets the different absorption properties of type-1 and type-2 AGN as an orientation effect. The heavy neutral absorption observed in type-2 AGN is attributed to the putative dusty torus that intercepts our line of view. On the other hand, the relatively unabsorbed soft X-ray spectrum seen in type-1 AGN is usually interpreted as direct vision of the innermost region of the AGN. This is also in agreement with the fact that ionized absorbers are mainly detected in unobscured AGN, since they are believed to be generated somewhere within the dusty torus (e.g. Krolik \& Kriss \cite{KK01}; Blustin et al. \cite{Blu05}), and therefore are extremely difficult to be detected in type-2 AGN. Since the unified models have successfully explained most of the observational properties of AGN, sources as NGC 7314, where both neutral and ionized absorbers can be detected at the same time, are the key to further test these models.

A possible explanation to the observational properties of NGC 7314 is that we are observing the central engine of the AGN, which accounts for the very rapid variability found in the lightcurve of the source and the broad component in the \FeKa{} line (Yaqoob et al \cite{Yaq96}, \cite{Yaq03}), but our line of sight grazes the torus. There is also increasing evidence that the torus is likely to be clumpy (e.g. Elitzur \cite{Eli06}; Nenkova et al. \cite{Nen08a}, \cite{Nen08b}; see also Jaffe et al. \cite{Jaf04}; Burtscher et al. \cite{Bur09}). If this is case, it could be possible that a line of sight that grazes the torus could allow to measure the ionized gas located in inner regions, as we observe with \XMM{}. This would be in agreement with the presence of \FeKa{} and \FeXXVI{} emission lines, originated by reflection in neutral and ionized matter, respectively. With the low statistics of the data, particularly with the RGS, we are not able to determine whether the WA is outflowing, or further constrain its location. For the purpose of describing this scenario we will assume that the ionized absorber is located within the locus of the torus, either originated as a wind in the inner edge of the torus or further in, since its actual location is not critical for our conclusions.

In this context, if some of the clouds of neutral gas that compose the torus intercept our line of sight, it would significantly obscure the inner parts of the torus thus making the ionized gas more elusive. This would be consistent with what we see in the \suz{} spectrum, which shows an increase in the neutral column density and the \FeKa{} line intensity, while the \FeXXVI{} line and WA are statistically undetected. This could also be the situation at the time of the \ASCA{} observations. However, the presence of a broad wing in the \FeKa{} profile and the very rapid variability at short timescales, suggest that we may be looking further in. Furthermore, as discussed in Sect.~\ref{asca}, \ASCA{} could in principle detect a WA with an $N_{\rm H}$ comparable to the one measured with \XMM{} if it was present (George at al. \cite{Geo98}). It has been proposed (Elvis et al. \cite{Elvis04}; Risaliti et al. \cite{Ris07}) that intervening clouds located in the BLR may be responsible for obscuration of the central source. Indeed, if the ionized absorber is originated in disk winds it could be hidden both by BLR and torus clouds. However, if that is the case the ionized gas must be outflowing at very high velocities (likely of the order of tens of thousands of km~s$^{-1}$), and it could be easily detected even with the low counts in the RGS spectrum. Therefore we conclude that the outflow velocities must be very modest, of the order of hundreds of km~s$^{-1}$, and likely associated to thermally driven winds from the torus.

A sketch of this possible scenario is shown in Fig.~\ref{sketch}. For illustrative purposes, the different parts of the sketch are not shown to scale. The left panel of Fig.~\ref{sketch} shows the situation at the moment of the \XMM{} observation. Our line of sight (solid line) is grazing the molecular clumpy torus and thus allows us to probe the inner regions of the AGN, including the ionized WA which is assumed to be located somewhere between the accretion disk and the inner edge of the torus. In this way both neutral and ionized absorbers can be detected, as well as emission lines of \FeKa{} and and \FeXXVI{} originated by reflection in distant matter (dashed lines). Likewise, the right panel of Fig.~\ref{sketch} shows the situation at the time of the \suz{} observation (and, possibly, also at the time of the \ASCA{} observations). Here, a cloud (or clouds) of gas from the torus intercepts our line of sight thus obscuring the inner parts of the AGN. This translates to an increase of the neutral absorbing column density, therefore making the detection of ionized matter along our line of sight much more challenging. Similarly, any emission lines of ionized matter would also be extremely difficult to detect. The increase in neutral matter would then favor the detection of the \FeKb{} line in addition to the \FeKa{} line.

We can estimate the time scales at which such changes in the absorption properties occur if they are caused by a single cloud crossing our line of sight. The cloud is assumed to be spherical with dimension $D \simeq N_{\rm H}/n$, where $N_{\rm H}$ and $n$ are the column density and density of the cloud, respectively. The time required for the cloud to pass transverse to our line of sight would be $t=D/v$, where $v=\sqrt{GM/R}$ is the velocity of the cloud assuming Keplerian motion. The mass of the black hole is $M \sim 5\times 10^6$~M$_{\odot}$ (Padovani \& Rafanelli \cite{PR88}). The distance from the cloud to the central black hole $R$ is difficult to estimate for the reasons exposed above, but we can assume the location of the inner edge of the torus as an upper limit. Using the approximate relation of Krolik \& Kriss (\cite{KK01}), $R\sim L_{44}^{1/2}$~pc, where $L_{44}$ is the ionizing luminosity of the source in units of $10^{44}$~erg~s$^{-1}$. In this work we measure an ionizing luminosity of $\sim 3.1 \times 10^{43}$~erg~s$^{-1}$ which gives an upper limit for the distance of $R \sim 0.6$~pc. At this distance the cloud would be moving at $v \sim 200$~km~s$^{-1}$. We also assume that the dimension of the cloud is large enough to totally obscure the ionizing source so that $D \gtrsim 10r_{\rm S}$, where $r_{\rm S}$ is the Schwarzschild radius of the black hole. Therefore, a cloud with $D \sim 1.5 \times 10^{13}$~cm would require at least $t > D/v \sim 9$~days to completely cross our line of sight. Of course, if the cloud is located at lower $R$ its velocity would increase and it would complete the transit faster. In either case the time span between observations are much higher than this value ($\sim$5~years between the last \ASCA{} and \XMM{} observations, and $\sim$6~years between the \XMM{} and \suz{} observations), which may suggest a large number of clouds. The observations available are much shorter than $t$, so that a transit cannot be monitored within a single observation. Further monitoring and additional and deeper observations of this source are therefore required to further test this scenario, as NGC 7314 seems to be an ideal laboratory to test AGN unification models.

%______________________________________________________________

\section{Conclusions}
\label{conclusions}

We have analyzed the archival \ASCA{}, \XMM{}, and \suz{} X-ray spectra of the NLS1/Sy 1.9 galaxy NGC 7314, taken between 1994 and 2007.

The EPIC-pn spectrum is modeled by a power-law plus a reflection component corrected from foreground Galactic absorption and neutral absorption intrinsic to the source. We also find significant evidence of ionized gas in the EPIC-pn spectrum, which may present up to three significantly distinct ionization phases. The RGS spectrum also show evidence of at least two phases of this WA, although with less significance due to the low statistics. At higher energies, we detect a \FeKa{} emission line at $6.40 \pm 0.03$~keV and an excess at $6.91 \pm 0.03$~keV that we identify with \FeXXVI{} in emission.

The \suz{} XIS and \ASCA{} SIS spectra, on the other hand, do not reveal significant ionized absorption but almost three times more neutral absorption than in the \XMM{} observation. In the case of the XIS spectrum, part of this neutral absorption could be attributed to an unresolved low-ionization gas, although the absence of high-ionization gas is significant with respect to the \XMM{} observation. The \suz{} XIS spectrum also shows \FeKa{} in emission at $6.41 \pm 0.02$~keV, and a narrow emission feature at $7.05 \pm 0.02$~keV which we identify with \FeKb{}, although the unusually low \FeKa{}/\FeKb{} flux ratio may indicate some contamination from a statistically undetected \FeXXVI{} line.

We interpret these observations in the context of the unified models of AGN as variations in the absorption properties of NGC 7314. If the torus were clumpy (e.g. Elitzur \cite{Eli06}; Nenkova et al. \cite{Nen08a}, \cite{Nen08b}) the differences between the \XMM{}, and \suz{} and \ASCA{} observations might be explained in terms of a cloud of neutral gas crossing our line of sight, which would be grazing the edge of the torus. In that case we would see an increase in the neutral column density and the suppression of the ionized gas signatures in the spectrum, as the WA is assumed to be located within the torus. Given the time span between the observation, and the short exposure times of the archival observations, additional monitoring and observations of NGC 7314 are needed to further test this scenario and, eventually, constrain location and kinematic properties of the ionized gas.

\begin{acknowledgements}
We would like to thank Hiroki Akamatsu for helping with the \suz{} data reduction. The Space Research Organization of The Netherlands is supported financially by NWO, the Netherlands Organization for Scientific Research. This research has made use of the Tartarus (Version 3.1) database, created by Paul O'Neill and Kirpal Nandra at Imperial College London, and Jane Turner at NASA/GSFC. Tartarus is supported by funding from PPARC, and NASA grants NAG5-7385 and NAG5-7067. We thank the anonymous referee for useful comments that improved this paper.
\end{acknowledgements}

\end{document}